\documentclass[aps,prd,amsmath,amssymb,latexsym,array,enumerate,letter,twocolumn,superscriptaddress,showpacs]{revtex4-1}
\pdfoutput=1
\usepackage{amsthm} 
\usepackage{bm} 
\pdfoutput=1
\allowdisplaybreaks
\usepackage{graphicx,color}
\usepackage[colorlinks=true,citecolor=blue,linkcolor=blue,urlcolor=blue]{hyperref}
\usepackage{caption}
\usepackage{subcaption}
\usepackage{footmisc}
\usepackage{bm}

 \usepackage{float}
\restylefloat{table}
\hypersetup{
    colorlinks,
    citecolor=black,
    filecolor=black,
    linkcolor=blue,
    urlcolor=blue
}
\def\be{\begin{equation}}
\def\ee{\end{equation}}
\def\bea{\begin{eqnarray}}
\def\eea{\end{eqnarray}}

\def\ms{M_\odot}

\usepackage{setspace}

\begin{document}

\title{Probing up-down quark matter via gravitational waves}
\author{Chen Zhang}
\email{czhang@physics.utoronto.ca}
\vskip 0.5cm

\affiliation{Department of Physics, University of Toronto, 60 St. George Street, Toronto, Ontario, Canada M5S 1A7}
\begin{abstract}
Recently, it was shown that quark matter with only $u$ and $d$ quarks ($ud$QM) can be the ground state of matter for baryon numbers $A>A_\textrm{min}$ with $A_{\rm min}\gtrsim 300$. In this paper, we explore $ud$ quark stars ($ud$QSs) that are composed of $ud$QM, in the context of the two-families scenario in which $ud$QSs and hadronic stars (HSs) can coexist. Distinct signatures are discussed compared to the conventional study regarding strange quark stars (SQSs). We show that the requirements of $A_{\rm min}\gtrsim 300$ and the most massive compact star observed being a $ud$QS together may put stringent constraints on the allowed parameter space of $ud$QSs. Then, we study the related gravitational-wave probe of the tidal deformability in binary star mergers, including the $ud$QS-$ud$QS and $ud$QS-HS cases. The obtained values of the tidal deformability at 1.4 solar masses and the average tidal deformability are all in good compatibility with the experimental constraints of GW170817. This study points to a new possible interpretation of the GW170817 binary merger event, where $ud$QS may be at least one component of the binary system detected.
\end{abstract}
\maketitle
\section{Introduction}
In the conventional picture of nuclear physics, quarks are confined in the state of hadrons. However, it is also possible that quark matter, a state consisting of deconfined quarks, exists. Bodmer~\cite{Bodmer:1971we}, Witten~\cite{Witten} and Terazawa~\cite{Terazawa:1979hq} proposed the hypothesis that quark matter with comparable numbers of $u$, $d$, and $s$ quarks, also called strange quark matter (SQM), might be the ground state of baryonic matter. However, this hypothesis is based on the bag model that cannot adequately model the flavor-dependent feedback of the quark gas on the QCD vacuum. Improved models have shown that quark matter with only $u$, and $d$ quarks ($ud$QM) is more stable than SQM~\cite{Buballa:1998pr,Wang:2002pza,HRZ2017,Wang:2019gam}, but with the common conclusion that neither is more stable than ordinary nuclei. In a recent study~\cite{HRZ2017}, with a phenomenological quark-meson model that can give good fits to all the masses and decay widths of the light meson nonets and can account for the flavor-dependent feedback~\cite{Osipov1,Osipov2}, the authors demonstrated that $ud$QM can be more stable than the ordinary nuclear matter and SQM when the baryon number $A$ is sufficiently large above $A_{\rm min}\gtrsim 300$. The absolute stability of $ud$QM is tested to be robust within $10\%$ departures of the experimental data.
The large $A_{\rm min}$ ensures the stability of ordinary nuclei in the periodic table, which also results in a large positive charge. Recently, a collider search for such high-electric-charge objects was attempted using LHC data~\cite{Aad:2019pfm}.  

One can also look for the evidence of $ud$QM from gravitational-wave detection experiments. The binary merger of compact stars produces strong gravitational wave fields, the waveforms of which encode the information of the tidal deformation that is sensitive to the matter equation of state (EOS). In general, stars with stiff EOSs can be tidally deformed easily due to their large radii. 

The GW170817 event detected by LIGO~\cite{GW170817} is the first confirmed merger event of compact stars. Together with the subsequent detection of the electromagnetic counterpart, GRB 170817A and  AT2017gfo~\cite{Abbott2}, they inspired a lot of studies that greatly move our understanding of nuclear matter forward~\cite{Abbott:2018exr, Radice:2017lry,Bauswein:2019skm,Kiuchi:2019lls,Annala:2017llu,Fattoyev:2017jql,De:2018uhw, Burgio:2018yix, Most:2018hfd, Most:2018eaw,Weih:2019rzo,Montana:2018bkb,Dexheimer:2019pay, Drago:2017bnf}. The initial analysis~\cite{GW170817} determines the chirp mass of the binary to be $M_c = 1.188\,\ms$. For the low-spin case, the binary mass ratio $q=M_2/M_1$ is constrained to the range $q=$0.7--1.0. Upper bounds have been placed on the tidal deformability at 1.4 solar masses $ \Lambda (1.4\, \ms) \lesssim 800$, and on the average tidal deformability $ \tilde{\Lambda} \leq 800$ at a $90\%$ confidence level. Later, an improved LIGO analysis~\cite{Abbott:2018wiz} gives $M_c=1.186^{+0.001}_{-0.001} \, M_{\odot}$, and a $90\%$ highest posterior density interval of $ \tilde{\Lambda}=300^{+420}_{-230}$ with $q=$0.73--1.00 for the low-spin prior. Lower bounds have been placed from AT2017gfo with kilonova models~\cite{Radice:2017lry,Kiuchi:2019lls, Bauswein:2019skm}. However, to the author's knowledge, the more strict lower bounds obtained in such analysis, including $\Lambda (1.4\, \ms) \gtrsim 200$~\cite{Bauswein:2019skm} and  $\tilde{\Lambda} \gtrsim 242$~\cite{Kiuchi:2019lls}, are all assuming the neutron star EOS. Therefore, we will not use them to constrain our study of quark stars here.

Conventionally, binary mergers are studied in the one-family scenario where it is assumed that all compact stars are within one family of hadronic matter EOSs~\cite{GW170817, Annala:2017llu,Fattoyev:2017jql,De:2018uhw,Burgio:2018yix}. However, the discovery of pulsars with large masses above $2\, M_\odot$ ~\cite{Demorest:2010bx,Antoniadis:2013pzd,Cromartie:2019kug} ruled out a large number of soft EOSs that were expected with the presence of hyperons and $\Delta$ resonances in the interiors. Therefore, it is natural to expect that the stars with masses above $2\, M_\odot$ and large radii are actually quark stars (QSs), and most of the ones with small masses and small radii are the hadronic stars (HSs). This possibility is the so-called ``two-families" scenario, which is based on the hypothesis that absolutely stable quark matter (either SQM or $ud$QM) exists, and that the hadronic stars can coexist with quark stars~\cite{Drago:2013fsa}.  The binary merger in the two-families scenario includes three cases: HS-HS~\cite{Drago:2017bnf}, HS-QS~\cite{Burgio:2018yix}, and QS-QS~\cite{Zhou:2017pha}. Alternatively, dropping the hypothesis that quark matter is the ground state gives the twin-stars scenario~\cite{Burgio:2018yix,Montana:2018bkb,Dexheimer:2019pay}, where quark matter only appears in the interiors of hybrid stars. 
 
Several things make $ud$ quark stars ($ud$QSs), which are composed of $ud$QM, very distinct compared to the strange quark stars (SQSs) that are composed of SQM. First, $ud$QSs can satisfy the $2\, M_\odot$ constraint more easily than HSs and SQSs~\cite{HRZ2017,Zhao:2019xqy} due to the nonhyperonic composition and the small effective bag constant.  Second, the coexistence of HSs and QSs requires that the conversion of hadronic matter to quark matter be neither too fast nor too slow compared to the age of our Universe. In contrast to the coexistence study for SQSs, where the conversion requires the presence of hyperons which only emerge above 1.5 solar masses, the conversion regarding $ud$QSs can happen at a smaller mass range since no hyperonic composition is needed. Therefore, it is possible that $ud$QSs can coexist with HSs even at the small mass range below $1.5 \, M_{\odot}$. This reasoning raises the possibility for GW170817 being a $ud$QS-$ud$QS merger or a $ud$QS-HS merger despite the smallness of the chirp mass $1.186 \, M_{\odot}$ and the high mass ratio $q=$0.73--1.00. Besides, the possibility of the QS-QS case is sometimes disfavored for GW170817/AT2017gfo because of the kilonova observation of nuclear radioactive decay~\cite{Drago:2018nzf}. However, it is possible that the $ud$QM ejected is quickly destabilized by the finite-size effects and converts into ordinary or heavy nuclei. The conversion is far more rapid for $ud$QM than for SQM, due to a much larger $A_{\rm min}$ and the nonstrange composition, so that there is no need to involve extra weak interactions to convert away strangeness. Note that the radii constraints derived from GW170817 are mostly for hadronic EOSs in the context of the one-family scenario~\cite{Annala:2017llu,Fattoyev:2017jql,De:2018uhw}, so that they have not much relevance to the $ud$QSs in the two-families scenario we are discussing here.

Motivated by these considerations, we explore the properties of $ud$QSs and the related gravitational-wave probe in the two-families scenario, including the binary merger cases $ud$QS-$ud$QS and $ud$QS-HS. We will discuss the related compatibility and constraints from GW170817. Note that we ignore the discussion of the HS-HS case since this possibility is not directly related to the study of quark stars and is disfavored to some extent for GW170817 based on the consideration of prompt collapse~\cite{Drago:2017bnf}.

\section{Properties of $ud$QS\lowercase{s}}
The EOS of $ud$QM can be well approximated by the simple form $p=1/3\,(\rho-\rho_s)$, where $\rho_s$ is the finite density at the surface. For the EOS of SQM, the coefficient $1/3$ is modified by the strange quark mass effect, with the $\rho_s$ value also being different. In the region of interest for $ud$QM, we can take the relativistic limit where the energy per baryon number in the bulk limit takes the form $E/A=\rho/n_A\approx (\chi N_C p_F^4/4\pi^2+B_{\rm eff})/n_A =3/4\, N_C p_F \chi+3\pi^2 B_{\rm eff}/p_F^3$~\cite{HRZ2017}, where $N_C=3$ is the color factor and $\chi=\sum_i f_i^{4/3}$ is the flavor factor, with the fraction $f_u=1/3=1/2\,f_d$ for $ud$QM. The effective Fermi momentum is $p_F=(3\pi^2 n_A)^{1/3}$. $B_{\rm eff}$ is the effective bag constant that accounts for the QCD vacuum contribution. Note that in this $ud$QM study, we can approximate $B_{\rm eff}$ as an effective constant since its dependence on flavor and density only causes a substantial effect when strangeness turns on at very large density~\cite{HRZ2017,Buballa:1998pr,Wang:2002pza}. Minimizing the energy per baryon number with respect to $p_F$ for fixed flavor composition gives
\be
\frac{E}{A}=3\sqrt{2\pi}\left(\chi^3 B_{\rm eff}\right)^{1/4}, 
\label{EperA_analy}
\ee
at which $p=0$, $\rho=\rho_s=4 B_{\rm eff}$. Equation~(\ref{EperA_analy}) matches the exact numerical result of the phenomenological meson model~\cite{HRZ2017} extremely well, with a mere error $\sim0.3\%$ due to a tiny $u$ $(d)$ quark mass. It was shown in Refs.~\cite{HRZ2017,Buballa:1998pr,Wang:2002pza} that $B_{\rm eff}$ has a smaller value in the two-flavor case than in the three-flavor case, so that $ud$QM is more stable than $\rm SQM$ in the bulk limit. Absolute stability of $ud$QM in the bulk limit implies $ E/A\lesssim 930 \,\rm MeV$, which corresponds to
\be
B_{\rm eff}\lesssim56.8 \, \rm MeV/fm^3
\label{B_stab}
\ee
from Eq.~(\ref{EperA_analy}). In general, a larger $E/A$ or $B_{\rm eff}$ gives a larger $A_{\rm min}$. The stability of ordinary nuclei against $ud$QM requires $A_{\rm min}\gtrsim300$, which translates to $E/A \gtrsim 903 \, \rm MeV$ or
\be
 B_{\rm eff}\gtrsim 50 \rm \, MeV/fm^3 
 \label{B_lower}
\ee
for the quark-meson model that matches the low-energy phenomenology~\cite{HRZ2017}. This quark-meson model also results in a quark-vacuum surface tension $\sigma\approx (91\,\rm MeV)^3$ that is robust against parameter variations.  

\begin{figure}[h]
 \centering
\includegraphics[width=8cm]{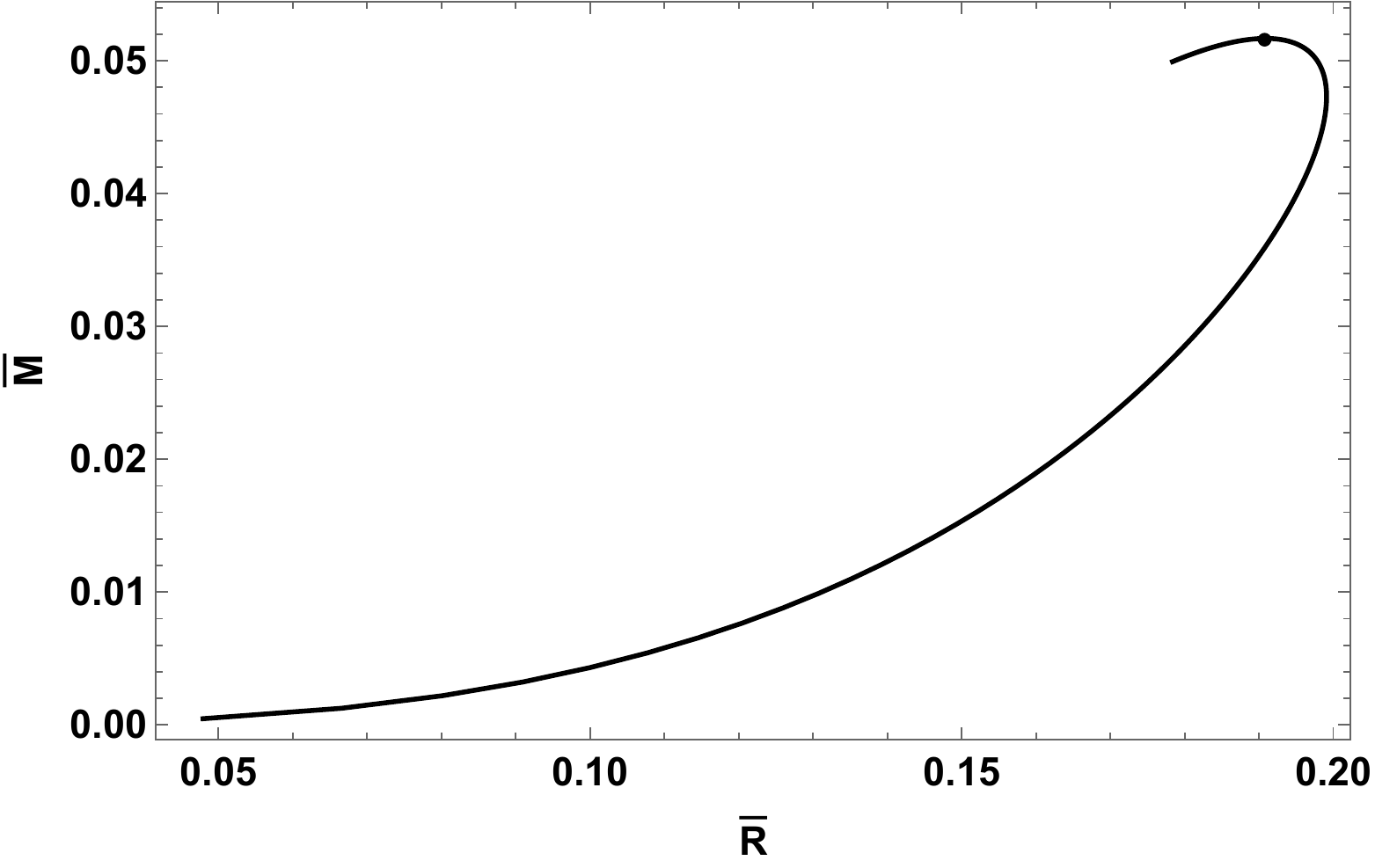}  
\caption{$\bar{M}$-$\bar{R}$ of $ud$QSs. The black dot at $(\bar{M},\bar{R})$= $(0.0517, 0.191)$ denotes the maximum mass configuration.}
   \label{fig:MR_dimless}
\end{figure}

\begin{figure}[h]
 \centering
\includegraphics[width=8cm]{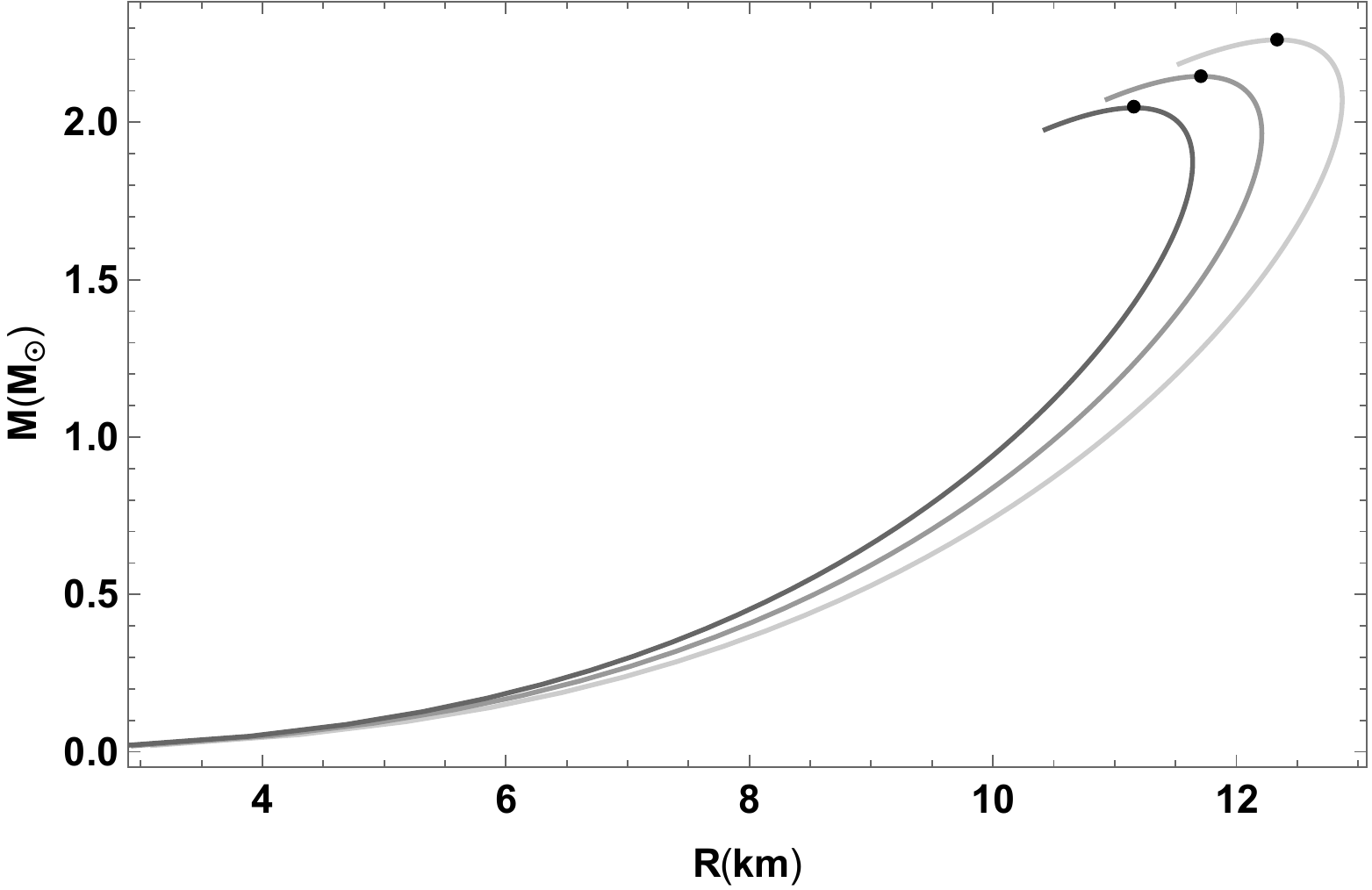}  
\caption{$M$-$R$ of $ud$QSs. Lines with darker color denote a larger effective bag constant $B_{\rm eff}$ and sample $(45, 50, 55)\rm\, MeV/fm^3$ respectively. The black dots denote the maximum mass location.}
   \label{fig:MR_B}
\end{figure}

The linear feature of $ud$QM EOS makes it possible to perform a dimensionless rescaling on parameters~\cite{Zdunik:2000xx,Haensel:2007yy}
\be
\bar{\rho}=\frac{\rho}{4\,B_{\rm eff}}, \,\, \bar{p}=\frac{p}{4\,B_{\rm eff}},  \,\,  \bar{r}={r}{\sqrt{4\,B_{\rm eff}}}, \,\, \bar{m}=m{\sqrt{4\,B_{\rm eff}}},
\label{eq:scaling}
\ee
 which enter the Tolman–Oppenheimer–Volkoff (TOV) equation~\cite{Oppenheimer:1939ne,Tolman:1939jz}
 \bea
 \begin{aligned}
{dp(r)\over dr}&=-{\left[m(r)+4\pi r^3p(r)\right]\left[\rho(r)+p(r)\right]\over r(r-2m(r))}\,,\,\,\\
{dm(r)\over dr}&=4\pi\rho(r)r^2,\, 
\end{aligned}
\label{eq:tov}
\eea
so that the rescaled solution is also dimensionless, and thus is independent of any specific value of $B_{\rm eff}$. The TOV solution with a specific $B_{\rm eff}$ value can be obtained directly from rescaling the dimensionless solution back with Eq.~(\ref{eq:scaling}). Solving the rescaled TOV equation with the $ud$QM EOS gives the dimensionless result shown in Fig.~\ref{fig:MR_dimless}, with the maximum rescaled mass at $(\bar{M},\bar{R})= (M{\sqrt{4\,B_{\rm eff}}},R{\sqrt{4\,B_{\rm eff}}})=(0.0517, 0.191),$ mapping to $M_{\rm max}\approx 15.174/\sqrt{B_{\rm eff}} \,\,M_\odot$, $R_{\rm M\,max}\approx 82.79/\sqrt{B_{\rm eff}}  \rm\, \, km$. Therefore, the requirement that $ud$QSs have a maximum mass not smaller than the recently observed most massive compact star J0740+6620 ($M\approx 2.14^{+0.10}_{-0.09}\,M_{\odot}$)~\cite{Cromartie:2019kug} implies
\be
B_{\rm eff}\lesssim50.3^{+4.5}_{-4.4}\,\rm MeV/fm^3,
\label{B_max}
\ee
which constrains more strictly than what Eq.~(\ref{B_stab}) imposes. Interestingly, the central value of the upper bound Eq.~(\ref{B_max}) is very close to the lower bound Eq.~(\ref{B_lower}) at the critical value $B_{c}\approx\rm  50\, MeV/fm^3$. To be more conservative, we can take 10\% departures, considering the theoretical and experimental uncertainties~\cite{Berger:1986ps,Lugones:2013ema,Ke:2013wga,Garcia:2013eaa,Palhares:2010be,Pinto:2012aq,Fraga:2018cvr,Cromartie:2019kug}, so that the allowed window of $B_{\rm eff}$ for $ud$QS is
\bea
\begin{aligned} 
 \{ B_{ ud\rm QS} \} &\approx [45 , \,55] \rm \, MeV/fm^3
 \end{aligned}
\eea
with the central value $B_{c}\approx 50 \rm\, MeV/fm^3$. The corresponding $M$-$R$ solution is shown in Fig.~\ref{fig:MR_B}. 

Note that some SQS studies~\cite{Burgio:2018yix,Zhou:2017pha,Weissenborn:2011qu} exploited similar small $B_{\rm eff}$ values to have maximum masses above $2 \,M_{\odot}$, but the smallness is not natural considering the appearance of strangeness, and a large perturbative QCD (pQCD) effect or a color superconducting phase has to be included to guarantee the stability. 

\begin{figure}[h]
  \centering
       \includegraphics[width=8.6cm]{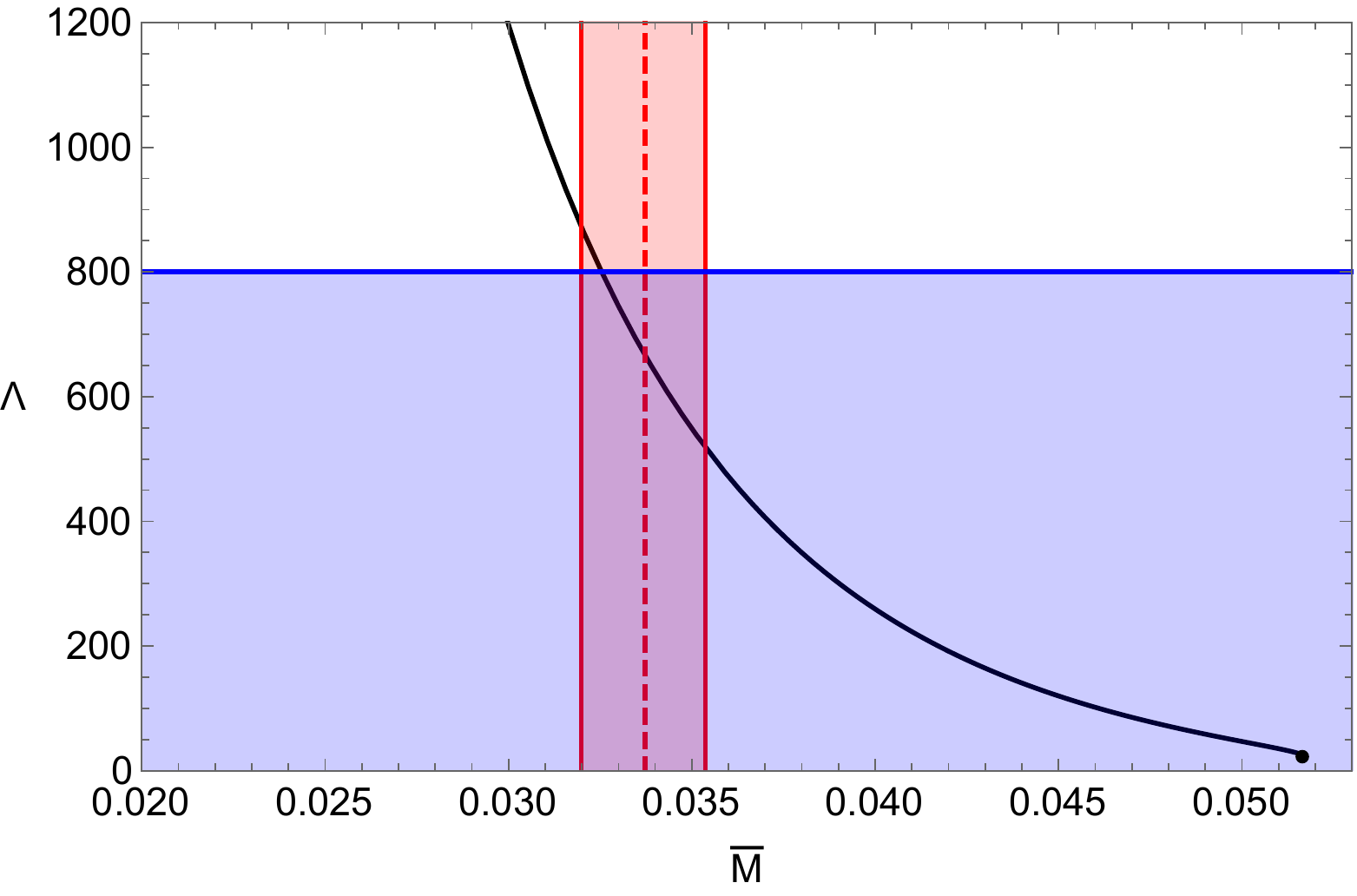}  
 \caption{The tidal deformability $\Lambda$ vs the rescaled star mass $\bar{M}$ for $ud$QSs. For $\bar{M}$ with $M=1.4 M_{\odot}$, the red band represents the region with $B_{\rm eff} \in \{ B_{ud\rm QS} \}$, and the red dashed line shows $B_{\rm eff} = B_{\rm c}$. The blue band denotes the GW170817 constraints on $\Lambda(1.4 \, M_{\odot})$~\cite{GW170817}.}
 \label{fig:LamvsM}
  \end{figure}

The response of compact stars to external disturbance is characterized by the Love number $k_2$~\cite{AELove,Hinderer:2007mb,Hinderer:2009ca,Postnikov:2010yn}, 
\bea
\begin{aligned}
k_2 &= \frac{8 C^5}{5} (1-2C)^2 [ 2 + 2 C (y_R-1) -y_R ] \\
&\times \{ 2 C [ 6- 3y_R + 3C (5y_R-8)]+4C^3[ 13-11 y_R \\&+ C (3y_R-2)+2C^2(1+y_R)]  \\
&+ 3 (1-2C)^2 [ 2 - y_R + 2C (y_R-1)] \log (1-2C )\}^{-1}~.
\label{eqn:k2}
\end{aligned}
\eea 
Here $C=M/R=C(\bar{M})$. And $y_R$ is $y(r)$ evaluated at the surface, which can be obtained by solving the following equation \cite{Postnikov:2010yn}:
\bea
\begin{aligned}
&ry^\prime(r)+y(r)^2+ r^2Q(r) \\
&+y(r)e^{\lambda(r)}\left[1+4\pi r^2(p(r)-\rho(r))\right]=0\,,
\label{eqn:y}
\end{aligned}
\eea
with the boundary condition $y(0)=2$. Here
\bea
\begin{aligned}
Q(r)&=4\pi e^{\lambda(r)} (5\rho(r)+9p(r)+\frac{\rho(r)+p(r)}{c_s^2(r)}) \\
&-6\frac{e^{\lambda(r)}}{r^2}-\left(\nu^\prime(r)\right)^2,
\label{eq:Q}
\end{aligned}
\eea
and 
\begin{equation}
e^{\lambda(r)}=\left[1-{2m(r)\over r}\right]^{-1}\,\,,\,
\nu^\prime(r)=2e^{\lambda(r)}{m(r)+4\pi p(r)r^3\over r^2}.
\label{eq:met}
\end{equation}
$c_s^2(r)\equiv dp/d\rho$ denotes the sound speed squared.
For stars with a finite surface density like quark stars, a matching condition should be used at the boundary $y_R^{\rm ext}=y_R^{\rm int}-  4\pi R^3\rho_s/M$~\cite{Damour:2009vw}. Solving Eq.~(\ref{eqn:y}) with the $\rho(r)$ and $p(r)$ obtained from Eq.~(\ref{eq:tov}), one obtains the function $k_2(C)$. The dimensionless tidal deformability $\Lambda=2k_2/(3C^5)$ as a function of mass $\bar{M}$ is thus obtained accordingly. 
The result is shown in Fig.~\ref{fig:LamvsM}.
  
  We see from Fig.~\ref{fig:LamvsM} that for $M=1.4\,M_\odot$ and $B_{\rm eff}\in \{ B_{ud\rm QS} \}$, one has $\bar{M}=M\sqrt{4\,B_{\rm eff}}\in [0.032, 0.035]$, as the red band in Fig.~\ref{fig:LamvsM} represents. Mapping this range to Fig.~\ref{fig:LamvsM} gives $\Lambda(1.4 \, M_\odot)\in [530, 857]$. And  $\Lambda(1.4\,M_\odot)\approx 670$ for $B_{\rm eff}=B_c$. We see that these results are well compatible with the GW170817 constraint $ \Lambda(1.4\,M_\odot)\lesssim 800$~\cite{GW170817}. In particular, the point where $\Lambda(1.4M_\odot)$ reaches the upper bound $\Lambda(1.4\,M_\odot)\sim 800$ sets a more stringent lower bound that $B_{ud\rm QS}\gtrsim47.9 \, \rm MeV/fm^3$. We also see that the result is not sensitive to the possible uncertainties related to the lower bound of $\Lambda(1.4\,M_\odot)$ constraint.
\section{Binary merger in the two-families scenario}
The average tidal deformability of a binary system is defined as
\bea
 \tilde \Lambda &=& \frac{16}{13} 
\frac{ (1+12q)}{(1+q)^5} {\Lambda} (M_1)+ \frac{16}{13}  \frac{q^4 (12+q)}{(1+q)^5}{\Lambda}({M}_2),
\label{LamLam}
\eea
where $M_1$ and $M_2$ are the masses of the binary components. And $q=M_2/M_1$, with $M_2$ being the smaller mass so that $0<q\leq1$. Then for any given chirp mass  $M_{c} = (M_1 M_2)^{3/5}/(M_1+M_2)^{1/5}$, one has
$M_2=(q^2(q+1))^{1/5} M_c \text{ and } M_1=((1+q)/q^3)^{1/5} M_c$. 
\subsection{$ud$QS-$ud$QS merger}
\begin{figure}[h]
  \centering
       \includegraphics[width=8.6cm]{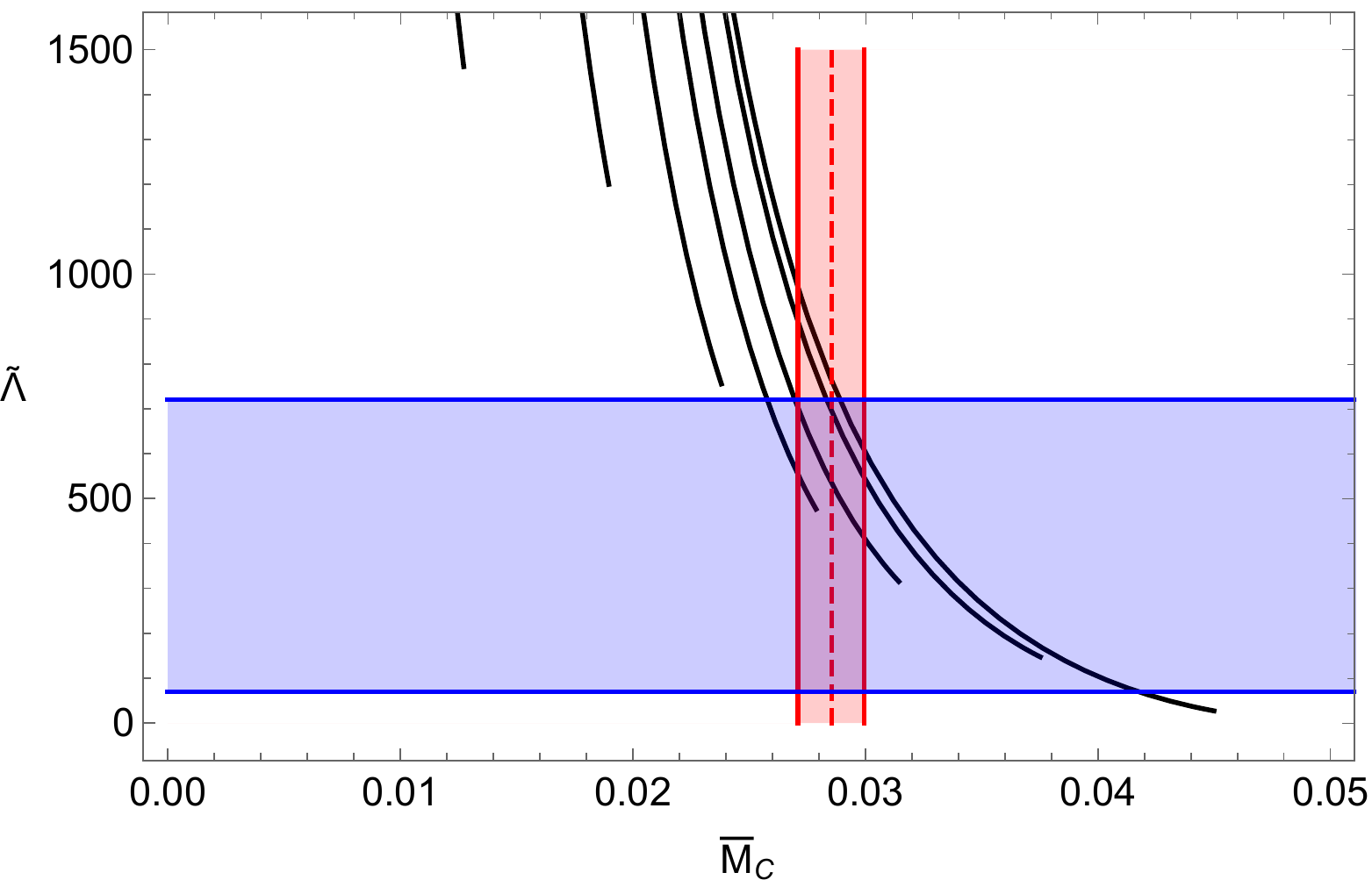}  
 \caption{The average tidal deformability $\tilde{\Lambda}$ vs the rescaled chirp mass $\bar{M}_c$ for the $ud$QS-$ud$QS merger case. Black curves show results with $q=M_2/M_1=(0.1, 0.2, 0.3,0.4, 0.5, 0.7, 1)$ from left to right, respectively. For the GW170817 event in which $M_c=1.186 \, M_\odot$, the red band represents the region of $\bar{M}_c$ with $B_{\rm eff}  \in \{ B_{ud\rm QS} \}$, and the red dashed line is with $B_{\rm eff}=B_{\rm c} $. The blue band is the GW170817 constraint on $\tilde{\Lambda}$~\cite{Abbott:2018wiz}.}
 \label{fig:LamvsMc}
  \end{figure}

In this case, the average tidal deformability can be expressed as a function of the rescaled mass parameter $\bar{M}=M\sqrt{4 B_{\rm eff}}$: 
\begin{equation}
 \tilde \Lambda= \frac{16}{13} 
\frac{ (1+12q)}{(1+q)^5} \Lambda (\bar{M}_1) + \frac{16}{13}  \frac{q^4 (12+q)}{(1+q)^5} \Lambda(\bar{M}_2).
\label{eq:LambdaQdim}
\end{equation} 
Substituting the $\Lambda(\bar{M})$ obtained previously into the formula above, we get the results shown in Fig.~\ref{fig:LamvsMc}.
  Note that in this figure, the lower end of each curve is determined by requiring each component of the binary system not to exceed its maximum mass. The $\bar{M}_c$ value of each end is negatively correlated with the $q$ value,  since for a given $M_c$, a less symmetric system has a larger component mass, which can exceed their maximum mass more easily.
The general shape of the figure matches our qualitative expectation. For given $\bar{M}_c$, a smaller mass ratio $q$ maps to a smaller $\tilde{\Lambda}$. Besides, for given $q$, a larger rescaled mass $\bar{M}_c=M_c \sqrt{4 B_{\rm eff}}$ corresponds to a smaller  $\tilde{\Lambda}$. These features are all due to the general fact that quark stars with larger masses have larger compactness, and thus are less likely to be tidally deformed. 

 As Fig.~\ref{fig:LamvsMc} shows, for GW170817 in which $M_c=1.186 \,M_\odot$, the constraint $\tilde{\Lambda}=300^{+420}_{-230}$~\cite{Abbott:2018wiz} translates to $0.4\lesssim q\lesssim 1$ for $B_{\rm eff} \in \{B_{ud\rm QS}\}$, and especially to $q=0.74$ for $B_{\rm eff} =B_c=50 \, \rm \, MeV/fm^3$, all of which are compatible with the GW170817 constraint $q=$0.73--1.00~\cite{GW170817}.  We see that $q\gtrsim 0.73$ and $\tilde{\Lambda}\lesssim720$ set a more stringent lower bound that $B_{ud\rm QS}\gtrsim 49.5\,\rm MeV/fm^3$. We also see that $B_{ud\rm QS}$ is not constrained much by the lower bound of $\tilde{\Lambda}$.
 
 \subsection{$ud$QS-HS merger}
\begin{figure}[h]
  \centering
       \includegraphics[width=8.6cm]{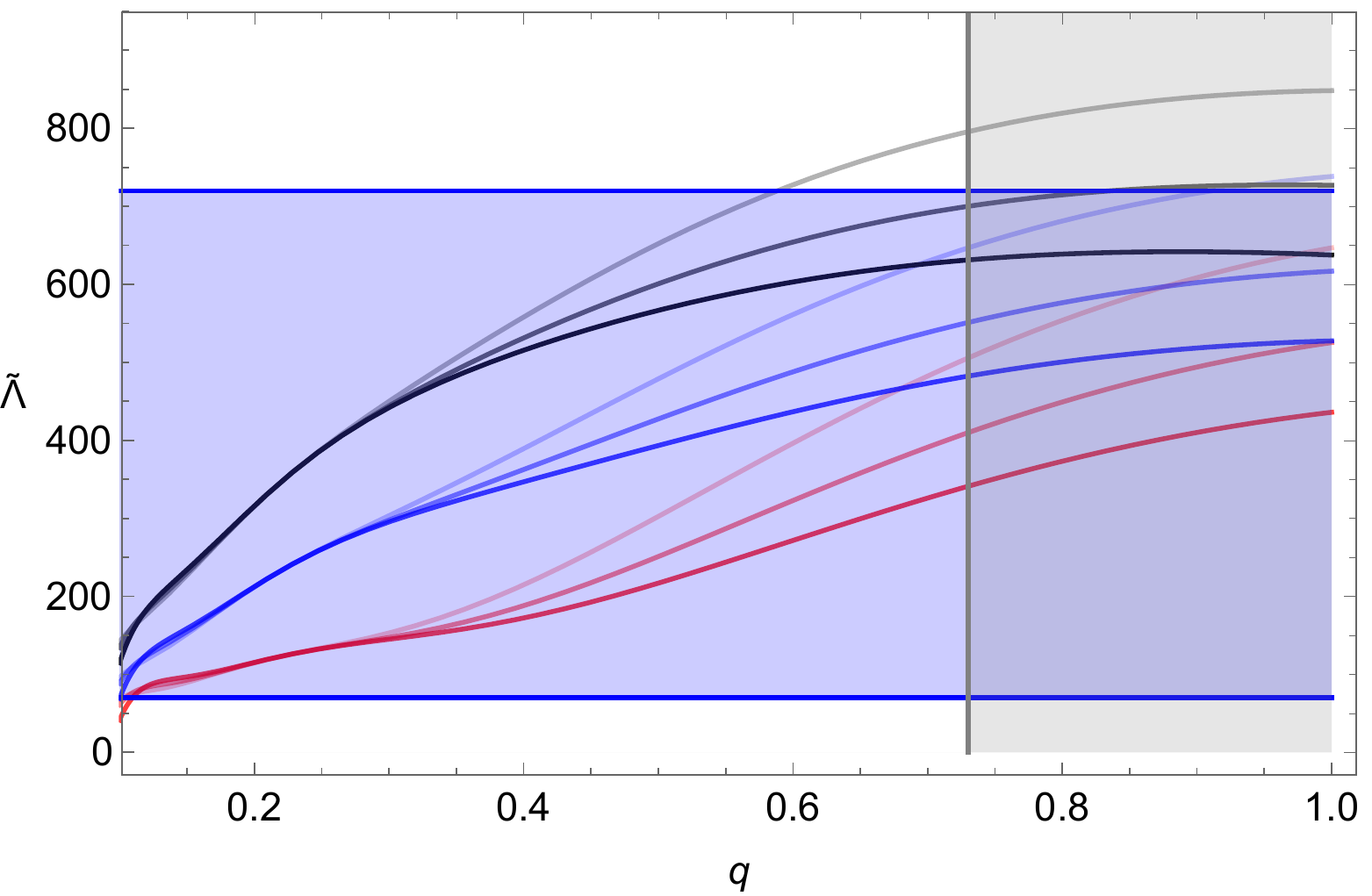} 
 \caption{The average tidal deformability $\tilde{\Lambda}$ vs $q=M_2/ M_1$ for the $ud$QS-HS merger case, with $M_2$ being the mass of the hadronic star and $M_{c}=1.186 \,M_{\odot}$ for the GW170817 event. For HS EOSs, SLy (blue), Bsk19 (red), and Bsk21 (black) are used. For $ud$QS EOSs, lines with darker color denote with larger $B_{\rm eff}$, sampling $(45, 50, 55){\rm\, MeV/fm^3}\in \{ B_{ud\rm QS} \}$, respectively. The blue band and gray band are the GW170817 constraints on $\tilde{\Lambda}$ and $q$, respectively~\cite{Abbott:2018wiz}.}
 \label{fig:QSHS}
  \end{figure}

For the $ud$QS-HS merger case, we need the information of the hadronic matter EOS, which has large uncertainties in the intermediate-density region. Based on nuclear physics alone, the EOS should match the low-density many-body calculation and high-density pQCD result~\cite{Oter:2019kig}. Here we use three benchmarks of hadron matter EOSs---SLy~\cite{Douchin:2001sv, Haensel:2004nu}, Bsk19, and Bsk21~\cite{Potekhin:2013qqa}--- that have unified representations from low density to high density.  Bsk19 is an example of soft EOSs. HSs with Bsk19 have maximum mass $M_{\rm max}=1.86\, M_\odot< 2 \,M_\odot$ and $R_{1.4 \,M_\odot}=10.74\rm\,km<11 \rm \,km$. The feature of small masses and small radii is preferred for the typical HSs branch of the two-families scenario. For illustration, we also show benchmarks of a hard EOS (Bsk21) with $M_{\rm max}=2.27 \,M_\odot$, $R_{1.4 \,M_\odot}=12.57\rm\,km$, and a moderate one (SLy) with $M_{\rm max}=2.05 \,M_\odot$, $R_{1.4 M_\odot}=11.3\rm\,km$. With Eq.~(\ref{LamLam}), the $\Lambda(M)$ results of $ud$QS, and the HS EOS benchmarks, we obtain the average tidal deformability  $\tilde{\Lambda}$ of the $ud$QS-HS system, as shown in Fig.~\ref{fig:QSHS}. 
  
 We see from Fig.~\ref{fig:QSHS} that the order of $\tilde{\Lambda}$ for different HS EOSs matches the expectation from the general rule that a HS with a stiffer EOS or a QS with a smaller effective bag constant has a larger radius, and thus has larger deformability. Lines with different hadronic EOSs tend to merge at lower $q$ as $\tilde{\Lambda}$ gets dominated by the contribution of large-mass quark stars. We see a good compatibility with current GW170817 constraint $\tilde{\Lambda}=300^{+420}_{-230}$ when $q=$0.73--1.00, except for an exclusion of the stiffest hadronic EOS (Bsk21).

\section{Conclusions}
\label{conclusion}
We have discussed the distinct properties that make $ud$QSs good candidates for the two-families scenario in which hadronic stars can coexist with quark stars. We have shown that the requirements of $A_{\rm min}\gtrsim 300$ and $M_{\rm max}\gtrsim 2.14 \, M_{\odot}$ together stringently constrain the effective bag constant of $ud$QSs to $B_{\rm eff} \approx  50 \rm  \, MeV/fm^3$. A $10\%$ relaxation that accounts for the possible uncertainties gives the conservative range $B_{ud\rm QS}  \in [45 , \,55] \rm \, MeV/fm^3$. Then we studied the related gravitational-wave probe of tidal deformability of binary star mergers including the $ud$QS-$ud$QS and $ud$QS-HS cases. For the $ud$QS-$ud$QS case, the upper bound of tidal deformability and the binary mass ratio of GW170817 further confine the allowed parameter space to $B_{ud\rm QS}  \in [49.5 , \,55] \rm \, MeV/fm^3$. Also, with the dimensionless rescaling method used, the analysis can be straightforwardly generalized to an arbitrary binary chirp mass and effective bag constant for current and future gravitational-wave events. The $ud$QS-HS case is also well compatible with the GW170817 constraints. These point to a new possibility that GW170817 can be identified as either a $ud$QS-$ud$QS merger or a $ud$QS-HS merger event.
\vspace{2.5ex}
\begin{acknowledgments}
This research is supported in part by the Natural Sciences and Engineering Research Council of Canada. I thank Bob Holdom, Jing Ren, and Luciano Rezzolla for helpful discussions. I also thank Randy S. Conklin and Shenglin Jing for proofreading the article.
\end{acknowledgments}
\vspace{1ex}
\noindent\textit{Note Added:} As we were finalizing this paper, we became aware of the recent Ref.~\cite{Wang:2019jze}. With a special version of the NJL model, Ref.~\cite{Wang:2019jze} has some discussions on the $\Lambda(1.4M_\odot)$ of nonstrange quark stars for the low-spin case of GW170817, and the authors also found that $\Lambda(1.4M_\odot)$ can match the experimental constraints in a certain parameter space. However, they neglected the study of the two-families scenario and the corresponding average tidal deformability $\tilde{\Lambda}$. Also, they only explored the parameter space in which $ud$QM is more stable than SQM, with the parameter space where $ud$QM is even more stable than nuclear matter remaining uncertain in their model.



\begin{thebibliography}{}
\bibitem{Bodmer:1971we} 
  A.~R.~Bodmer,
  Phys.\ Rev.\ D {\bf 4}, 1601 (1971).
\bibitem{Witten}
  E.~Witten,
  Phys.\ Rev.\ D {\bf 30}, 272 (1984).
\bibitem{Terazawa:1979hq} 
  H.~Terazawa,
  INS-Report-336 (INS, University of Tokyo, Tokyo) May, 1979.
\bibitem{Buballa:1998pr} 
  M.~Buballa and M.~Oertel,    			    
  Phys.\ Lett.\ B {\bf 457}, 261 (1999)
  [hep-ph/9810529].

\bibitem{Wang:2002pza} 
  P.~Wang, V.~E.~Lyubovitskij, T.~Gutsche and A.~Faessler,
  Phys.\ Rev.\ C {\bf 67}, 015210 (2003)
  [hep-ph/0205251].

\bibitem{HRZ2017} 
  Bob Holdom, Jing Ren and Chen Zhang,
  Phys.\ Rev.\ Lett.\  {\bf 120}, no. 22, 222001 (2018)
  [arXiv:1707.06610 [hep-ph]].

\bibitem{Wang:2019gam} 
  Q.~Wang, T.~Zhao and H.~S.~Zong,
  arXiv:1908.01325 [hep-ph].
  
\bibitem{Osipov1} 
  A.~A.~Osipov, B.~Hiller and A.~H.~Blin,
  Phys.\ Rev.\ D {\bf 88}, 054032 (2013)
  [arXiv:1309.2497 [hep-ph]].
\bibitem{Osipov2} 
  J.~Moreira, J.~Morais, B.~Hiller, A.~A.~Osipov and A.~H.~Blin,
  Phys.\ Rev.\ D {\bf 98}, no. 7, 074010 (2018)
  [arXiv:1806.00327 [hep-ph]].
  


\bibitem{Aad:2019pfm} 
  G.~Aad {\it et al.} [ATLAS Collaboration],
  Phys.\ Rev.\ Lett.\  {\bf 124}, no. 3, 031802 (2020)
  [arXiv:1905.10130 [hep-ex]].


  \bibitem{GW170817} 
  B.~P.~Abbott {\it et al.} [LIGO Scientific and Virgo Collaborations],
  Phys.\ Rev.\ Lett.\  {\bf 119}, no. 16, 161101 (2017)
  [arXiv:1710.05832 [gr-qc]].
   \bibitem{Abbott2} 
 Abbott, B. P., et al. 2017b, Astrophys. J., 848, L13; —. 2017c, Astrophys. J., 848, L12


\bibitem{Abbott:2018exr} 
  B.~P.~Abbott {\it et al.} [LIGO Scientific and Virgo Collaborations],
  Phys.\ Rev.\ Lett.\  {\bf 121}, no. 16, 161101 (2018)
  [arXiv:1805.11581 [gr-qc]].

   \bibitem{Radice:2017lry} 
  D.~Radice, A.~Perego, F.~Zappa and S.~Bernuzzi,
  Astrophys.\ J.\  {\bf 852}, no. 2, L29 (2018)
  [arXiv:1711.03647 [astro-ph.HE]].
  \bibitem{Bauswein:2019skm} 
  A.~Bauswein {\it et al.},
  AIP Conf.\ Proc.\  {\bf 2127}, no. 1, 020013 (2019)
  [arXiv:1904.01306 [astro-ph.HE]].
  
\bibitem{Kiuchi:2019lls} 
  K.~Kiuchi, K.~Kyutoku, M.~Shibata and K.~Taniguchi
  Astrophys.\ J.\  {bf 876}, no. 2, L31 (2019)
  [arXiv:1903.01466 [astro-ph.HE]].
 
\bibitem{Annala:2017llu} 
  E.~Annala, T.~Gorda, A.~Kurkela and A.~Vuorinen,
  Phys.\ Rev.\ Lett.\  {\bf 120}, no. 17, 172703 (2018)
  [arXiv:1711.02644 [astro-ph.HE]].
  
\bibitem{Fattoyev:2017jql} 
  F.~J.~Fattoyev, J.~Piekarewicz and C.~J.~Horowitz,
  Phys.\ Rev.\ Lett.\  {\bf 120}, no. 17, 172702 (2018)
  [arXiv:1711.06615 [nucl-th]].
  
\bibitem{De:2018uhw} 
  S.~De, D.~Finstad, J.~M.~Lattimer, D.~A.~Brown, E.~Berger and C.~M.~Biwer,
  Phys.\ Rev.\ Lett.\  {\bf 121}, no. 9, 091102 (2018)
  Erratum: [Phys.\ Rev.\ Lett.\  {\bf 121}, no. 25, 259902 (2018)]
  [arXiv:1804.08583 [astro-ph.HE]].
  
 \bibitem{Drago:2017bnf} 
  A.~Drago and G.~Pagliara,
  Astrophys.\ J.\  {\bf 852}, no. 2, L32 (2018)
  [arXiv:1710.02003 [astro-ph.HE]].
  
    \bibitem{Burgio:2018yix} 
  G.~F.~Burgio, A.~Drago, G.~Pagliara, H.~J.~Schulze and J.~B.~Wei,
  Astrophys.\ J.\  {\bf 860}, no. 2, 139 (2018)
  [arXiv:1803.09696 [astro-ph.HE]].
 
  
\bibitem{Most:2018hfd} 
  E.~R.~Most, L.~R.~Weih, L.~Rezzolla and J.~Schaffner-Bielich,
  Phys.\ Rev.\ Lett.\  {\bf 120}, no. 26, 261103 (2018)
  [arXiv:1803.00549 [gr-qc]].

\bibitem{Most:2018eaw} 
  E.~R.~Most, L.~J.~Papenfort, V.~Dexheimer, M.~Hanauske, S.~Schramm, H.~St\"ocker, and L.~Rezzolla,
  Phys.\ Rev.\ Lett.\  {\bf 122}, no. 6, 061101 (2019)
  [arXiv:1807.03684 [astro-ph.HE]].

\bibitem{Weih:2019rzo} 
  L.~R.~Weih, E.~R.~Most and L.~Rezzolla,
  Astrophys.\ J.\  {\bf 881}, 73 (2019)
  [arXiv:1905.04900 [astro-ph.HE]].
  
\bibitem{Montana:2018bkb} 
  G.~Montana, L.~Tolos, M.~Hanauske and L.~Rezzolla,
  Phys.\ Rev.\ D {\bf 99}, no. 10, 103009 (2019)
  [arXiv:1811.10929 [astro-ph.HE]].

\bibitem{Dexheimer:2019pay} 
  V.~Dexheimer, L.~T.~T.~Soethe, J.~Roark, R.~O.~Gomes, S.~O.~Kepler and S.~Schramm,
  Int.\ J.\ Mod.\ Phys.\ E {\bf 27}, no. 11, 1830008 (2018)
  [arXiv:1901.03252 [astro-ph.HE]].
  
  \bibitem{Abbott:2018wiz} 
  B.~P.~Abbott {\it et al.} [LIGO Scientific and Virgo Collaborations],
  Phys.\ Rev.\ X {\bf 9}, no. 1, 011001 (2019)
  [arXiv:1805.11579 [gr-qc]].
  
\bibitem{Demorest:2010bx} 
  P.~Demorest, T.~Pennucci, S.~Ransom, M.~Roberts and J.~Hessels,
  Nature {\bf 467}, 1081 (2010)
  [arXiv:1010.5788 [astro-ph.HE]].
\bibitem{Antoniadis:2013pzd} 
  J.~Antoniadis {\it et al.},
  Science {\bf 340}, 6131 (2013)
  [arXiv:1304.6875 [astro-ph.HE]].
\bibitem{Cromartie:2019kug} 
  H.~T.~Cromartie {\it et al.},
  arXiv:1904.06759 [astro-ph.HE].
  

    \bibitem{Drago:2013fsa} 
  A.~Drago, A.~Lavagno and G.~Pagliara,
  Phys.\ Rev.\ D {\bf 89}, no. 4, 043014 (2014)
  [arXiv:1309.7263 [nucl-th]].
  
  \bibitem{Zhou:2017pha} 
  E.~P.~Zhou, X.~Zhou and A.~Li,
  Phys.\ Rev.\ D {\bf 97}, no. 8, 083015 (2018)
  [arXiv:1711.04312 [astro-ph.HE]].
  
\bibitem{Zhao:2019xqy} 
  T.~Zhao, W.~Zheng, F.~Wang, C.~M.~Li, Y.~Yan, Y.~F.~Huang and H.~S.~Zong,
  Phys.\ Rev.\ D {\bf 100}, no. 4, 043018 (2019)
  [arXiv:1904.09744 [nucl-th]].
 
 
  \bibitem{Drago:2018nzf} 
  A.~Drago, G.~Pagliara, S.~B.~Popov, S.~Traversi and G.~Wiktorowicz,
  Universe {\bf 4}, no. 3, 50 (2018)
  [arXiv:1802.02495 [astro-ph.HE]].


  

\bibitem{Zdunik:2000xx} 
  J.~L.~Zdunik,
  Astron.\ Astrophys.\  {\bf 359}, 311 (2000)
  [astro-ph/0004375].
\bibitem{Haensel:2007yy} 
  P.~Haensel, A.~Y.~Potekhin and D.~G.~Yakovlev,
  Astrophys.\ Space Sci.\ Libr.\  {\bf 326}, pp.1 (2007).


\bibitem{Tolman:1939jz} 
  R.~C.~Tolman,
  Phys.\ Rev.\  {\bf 55}, 364 (1939).
  
\bibitem{Oppenheimer:1939ne} 
  J.~R.~Oppenheimer and G.~M.~Volkoff,
  Phys.\ Rev.\  {\bf 55}, 374 (1939).
\bibitem{Berger:1986ps} 
  M.~S.~Berger and R.~L.~Jaffe,
  Phys.\ Rev.\ C {\bf 35}, 213 (1987).
  
\bibitem{Lugones:2013ema} 
  G.~Lugones, A.~G.~Grunfeld and M.A. Ajmi,
  Phys.\ Rev.\ C {\bf 88}, no. 4, 045803 (2013)
  [arXiv:1308.1452 [hep-ph]].

\bibitem{Ke:2013wga} 
  W.~y.~Ke and Y.~x.~Liu,
  Phys.\ Rev.\ D {\bf 89}, no. 7, 074041 (2014)
  [arXiv:1312.2295 [hep-ph]].

\bibitem{Garcia:2013eaa} 
  A.~F.~Garcia and M.~B.~Pinto,
  Phys.\ Rev.\ C {\bf 88}, no. 2, 025207 (2013)
  [arXiv:1306.3090 [hep-ph]].
\bibitem{Palhares:2010be} 
  L.~F.~Palhares and E.~S.~Fraga,
  Phys.\ Rev.\ D {\bf 82}, 125018 (2010)
  [arXiv:1006.2357 [hep-ph]].
\bibitem{Pinto:2012aq} 
  M.~B.~Pinto, V.~Koch and J.~Randrup,
  Phys.\ Rev.\ C {\bf 86}, 025203 (2012)
  [arXiv:1207.5186 [hep-ph]].

  
\bibitem{Fraga:2018cvr} 
  E.~S.~Fraga, M.~Hippert and A.~Schmitt,
  Phys.\ Rev.\ D {\bf 99}, no. 1, 014046 (2019)
  [arXiv:1810.13226 [hep-ph]].


\bibitem{Weissenborn:2011qu} 
  S.~Weissenborn, I.~Sagert, G.~Pagliara, M.~Hempel and J.~Schaffner-Bielich,
  Astrophys.\ J.\  {\bf 740}, L14 (2011)
  [arXiv:1102.2869 [astro-ph.HE]].

  \bibitem{AELove} A. E. H. Love, Proc. R. Soc. A 82, 73 (1909).
    \bibitem{Hinderer:2007mb} 
  T.~Hinderer,
  Astrophys.\ J.\  {\bf 677}, 1216 (2008)
  [arXiv:0711.2420 [astro-ph]].
\bibitem{Hinderer:2009ca} 
  T.~Hinderer, B.~D.~Lackey, R.~N.~Lang and J.~S.~Read,
  Phys.\ Rev.\ D {\bf 81}, 123016 (2010)
  [arXiv:0911.3535 [astro-ph.HE]].

  \bibitem{Postnikov:2010yn} 
  S.~Postnikov, M.~Prakash and J.~M.~Lattimer,
  Phys.\ Rev.\ D {\bf 82}, 024016 (2010)
  [arXiv:1004.5098 [astro-ph.SR]].
  
  \bibitem{Damour:2009vw} 
  T.~Damour and A.~Nagar,
  Phys.\ Rev.\ D {\bf 80}, 084035 (2009)
  [arXiv:0906.0096 [gr-qc]].


\bibitem{Oter:2019kig} 
  E.~L.~Oter, A.~Windisch, F.~J.~Llanes-Estrada and M.~Alford,
  J.\ Phys.\ G {\bf 46}, no. 8, 084001 (2019)
  [arXiv:1901.05271 [gr-qc]].


\bibitem{Douchin:2001sv} 
  F.~Douchin and P.~Haensel,
  Astron.\ Astrophys.\  {\bf 380}, 151 (2001)
  [astro-ph/0111092].

  
\bibitem{Haensel:2004nu} 
  P.~Haensel and A.~Y.~Potekhin,
  Astron.\ Astrophys.\  {\bf 428}, 191 (2004)
  [astro-ph/0408324].


\bibitem{Potekhin:2013qqa} 
  A.~Y.~Potekhin, A.~F.~Fantina, N.~Chamel, J.~M.~Pearson and S.~Goriely,
  Astron.\ Astrophys.\  {\bf 560}, A48 (2013)
  [arXiv:1310.0049 [astro-ph.SR]].


\bibitem{Wang:2019jze} 
  Q.~Wang, C.~Shi and H.~S.~Zong,
  Phys.\ Rev.\ D {\bf 100}, no. 12, 123003 (2019)
  Erratum: [Phys.\ Rev.\ D {\bf 100}, no. 12, 129903 (2019)]
  [arXiv:1908.06558 [hep-ph]].

  
\end{thebibliography}
\end{document}